\documentclass[preprint,amsmath,amssymb,aps,prd,showpacs]{revtex4}
\usepackage{graphicx}

\begin{document}
\def\be{\begin{equation}}
\def\ee{\end{equation}}
\def\bea{\begin{eqnarray}}
\def\eea{\end{eqnarray}}


%
%

\title{MASS QUADRUPOLE AS A SOURCE OF NAKED SINGULARITIES}

\author{HERNANDO QUEVEDO \footnote{Instituto de Ciencias Nucleares,
Universidad Nacional Aut\'onoma de M\'exico}}

\address{Dipartimento di Fisica, Universit\`a di Roma "La Sapienza"\\ Piazzale Aldo Moro 5, I-00185 Roma, Italy;\\  
ICRANet, Piazza della Repubblica 10, I-65122 Pescara, Italy  \\    
quevedo@nucleares.unam.mx}


\begin{abstract}
We investigate the gravitational field of a 
static mass with quadrupole moment in empty space.  
It is shown that in general this
configuration is characterized by the presence of curvature singularities
without a surrounding event horizon. These naked singularities 
generate an effective field of repulsive gravity which, in turn, 
drastically changes the behavior of test particles. As a possible consequence, 
the accretion disk around a naked singularity presents a particular 
discontinuous structure. 
\end{abstract}

\keywords{Quadrupole moment; naked singularities; accretion disks.}
\maketitle

\section{Introduction}	
\label{sec:int}

According to the black hole uniqueness theorems, the most general black hole solution
in empty space is described by the Kerr metric \cite{kerr63}
\begin{eqnarray}
ds^2 & = &  {{r^2 - 2Mr + a^2}\over {r^2 + a^2\cos^2 \theta}}
(dt - a\sin^2\theta d\varphi)^2 - {{\sin^2 \theta}\over {r^2 + a^2\cos^2 \theta}}[(r^2 + a^2) d\varphi - adt]^2 
\nonumber \\ &&
  - {{r^2 + a^2\cos^2 \theta}\over {r^2 - 2Mr + a^2}}dr^2 -
(r^2 + a^2\cos^2\theta)d\theta^2 \ ,
\nonumber
\end{eqnarray}
which represents the exterior gravitational field of a rotating 
mass $M$ with specific angular momentum  $a=J/M$. A true curvature 
singularity is determined by the equation $r^2 + a^2\cos^2\theta =0$ 
and is interpreted as a ring singularity situated on the equatorial plane
$\theta = \pi/2$. The ring singularity is isolated from the exterior 
space by an event horizon situated on a sphere of radius 
$r_h = M + \sqrt{M^2-a^2}$. In the case $a^2>M^2$ no event horizon exists 
and the ring singularity becomes naked. Different studies \cite{def78,cal79,rud98} show, however,
that, in realistic situations where astrophysical objects are surrounded by accreting matter, a
Kerr naked singularity is an unstable configuration that 
rapidly decays into a black hole. Moreover, it now seems established that a 
gravitational collapse cannot in generic situations lead to the formation of a
final configuration resembling the Kerr solution with a naked singularity.
These results indicate that rotating naked singularities cannot be very common 
objects in nature. 

The above results seem to support the validity of the cosmic censorship hypothesis \cite{penrose}
according to which a physically realistic gravitational collapse, which evolves from a regular initial state, 
can never lead to the formation of a naked singularity; that is, 
all singularities formed as the result of such a
collapse should always be enclosed within an event horizon and
hence invisible to outside observers. Many attempts have been made 
to prove this conjecture with the same mathematical rigor used to show the inevitability of 
singularities in general relativity \cite{hawking}. So far, no general proof has been formulated.
Instead, particular scenarios of gravitation collapse have been investigated some of which indeed 
corroborate the correctness of the conjecture. Other studies studies \cite{naked}, however, 
indicate that under certain circumstances naked singularities can appear as a result of 
a realistic gravitational collapse. Indeed, it turns out that in an inhomogeneous collapse, 
there exists a critical degree of inhomogeneity below which black holes 
form. Naked singularities appear if the degree of inhomogeneity is higher than the critical value.
The speed of the collapse and the shape of the collapsing object are also factors that play an 
important role in the determination of the final state of the collapse. Naked singularities form 
more frequently if the collapse occurs very rapidly and the object is not exactly spherically
symmetric.

In view of this situation it seems reasonable to investigate the effects of naked singularities 
on the surrounding spacetime. This is the main aim of the present work. We investigate 
the simplest generalization of the Schwarzschild metric that contains a naked singularity. 
In fact, we show that, starting from the Schwarzschild metric,
 the Zipoy--Voorhees \cite{zip66,voor70} transformation can be used 
to generate a static axisymmetric spacetime which describes the field of a mass 
with a particular quadrupole moment. For any values of the quadrupole, the spacetime
is characterized by the presence of naked singularities situated at a finite distance
from the origin of coordinates. We argue that the effects of the quadrupole 
can be described by means of an effective potential with an effective mass that,
for certain values of the quadrupole parameter and coordinates, can become negative, 
giving rise to phenomena related to repulsive gravity. This is valid for any gravitational 
configuration with quadrupole moment. An example is shown where an accretion disk becomes 
discontinuous because of the presence of a naked singularity in its center.

\section{The simplest static spacetime with  quadrupole moment}
\label{sec:zv}
Static axisymmetric gravitational fields in empty space can be described 
by the Weyl line element \cite{solutions}
\be
ds^2 = e^{2\psi} dt^2 - e^{-2\psi}\left[e^{2\gamma}(d\rho^2+dz^2)+\rho^2 d\varphi^2\right] \ ,
\ee
where $\psi=\psi(\rho,z)$ and $\gamma=\gamma(\rho,z)$. The field equations $\psi_{\rho\rho} + \psi_{zz} + \psi_{\rho}/\rho =0$,
$\gamma_\rho = \rho(\psi_\rho^2-\psi_z^2)$, and $\gamma_z = 2\rho\psi_\rho\psi_z$ are invariant with respect to 
the Zipoy--Voorhees \cite{zip66,voor70} (ZV) transformation  $\psi\rightarrow \delta\psi$ and $\gamma\rightarrow \delta^2\gamma$,
where $\delta$ is a real constant parameter. For any particular solution, the ZV transformation generates 
a family of solutions that is parametrized by $\delta$. 

In terms of multipole moments, the simplest static 
solution contained in the Weyl class is the Schwarzschild metric which is the only one that possesses a mass monopole moment only. From a physical 
point of view, the next interesting solution must describe the exterior field of a mass with quadrupole moment. In this case, it is possible to find a large number of exact solutions with the same quadrupole \cite{quev10} that differ only in the set of higher multipoles. A common characteristic of solutions with quadrupole is that their explicit form is rather cumbersome, making them difficult to be handled analytically \cite{quev90}.
To meliorate this situation, we derive here an exact solution with quadrupole which can be written in a compact and simple form. To this end, we consider the Schwarzschild solution and apply a ZV transformation with $\delta = 1 + q$, The resulting metric can be written in spherical 
coordinates as 
\be
ds^2 = \left(1-\frac{2m}{r}\right)^{1+q} dt^2  
\nonumber
\ee
\be
- \left(1-\frac{2m}{r}\right)^{-q}\left[ \left(1+\frac{m^2\sin^2\theta}{r^2-2mr}\right)^{-q(2+q)} \left(\frac{dr^2}{1-\frac{2m}{r}}+ r^2d\theta^2\right) + r^2 \sin^2\theta d\varphi^2\right].
\label{zv}
\ee
This solution is axially symmetric and reduces to the spherically symmetric Schwarzschild metric only for $q\rightarrow 0$. 
It is asymptotically flat for any finite values of the parameters $m$ and $q$. Moreover, in the limiting case $m\rightarrow 0$
it can be shown that, independently of the value of $q$, there exists a coordinate transformation that 
transforms the resulting metric into the Minkowski soution. This last property is important from a physical point of view 
because it means that the parameter $q$ is related to a genuine mass distribution. To see this explicitly,
we calculate the multipole moments of the solution by using the invariant definition proposed by Geroch \cite{ger}. The 
lowest mass multipole moments $M_n$, $n=0,1,\ldots $ are given by
\be 
M_0= (1+q)m\ , \quad M_2 = -\frac{m^3}{3}q(1+q)(2+q)\ ,
\ee
whereas higher moments are proportional to $mq$ and can be 
completely rewritten in terms of $M_0$ and $M_2$. This means that the arbitrary parameters $m$ and $q$ determine the mass and quadrupole 
which are the only independent multipole moments of the solution. In the limiting case $q=0$ only the monopole $M_0=m$ 
survives, as in the Schwarzschild spacetime. In the limit $m=0$, with $q\neq 0$, all moments vanish identically, implying that 
no mass distribution is present and the spacetime must be flat. This is in accordance with the result mentioned above for the 
metric (\ref{zv}). Furthermore, notice that all odd multipole moments are zero because the solution possesses an additional 
reflection symmetry with respect to the equatorial plane. 

We conclude that the above metric describes the exterior gravitational 
field of a static deformed mass. The deformation is described by the quadrupole moment $M_2$ which is positive for a prolate mass 
distribution and negative for an oblate one. Notice that the condition $q>-1$ must be satisfied in order to avoid the appearance of
a negative total mass $M_0$. 

\begin{figure}
\includegraphics[scale=0.2]{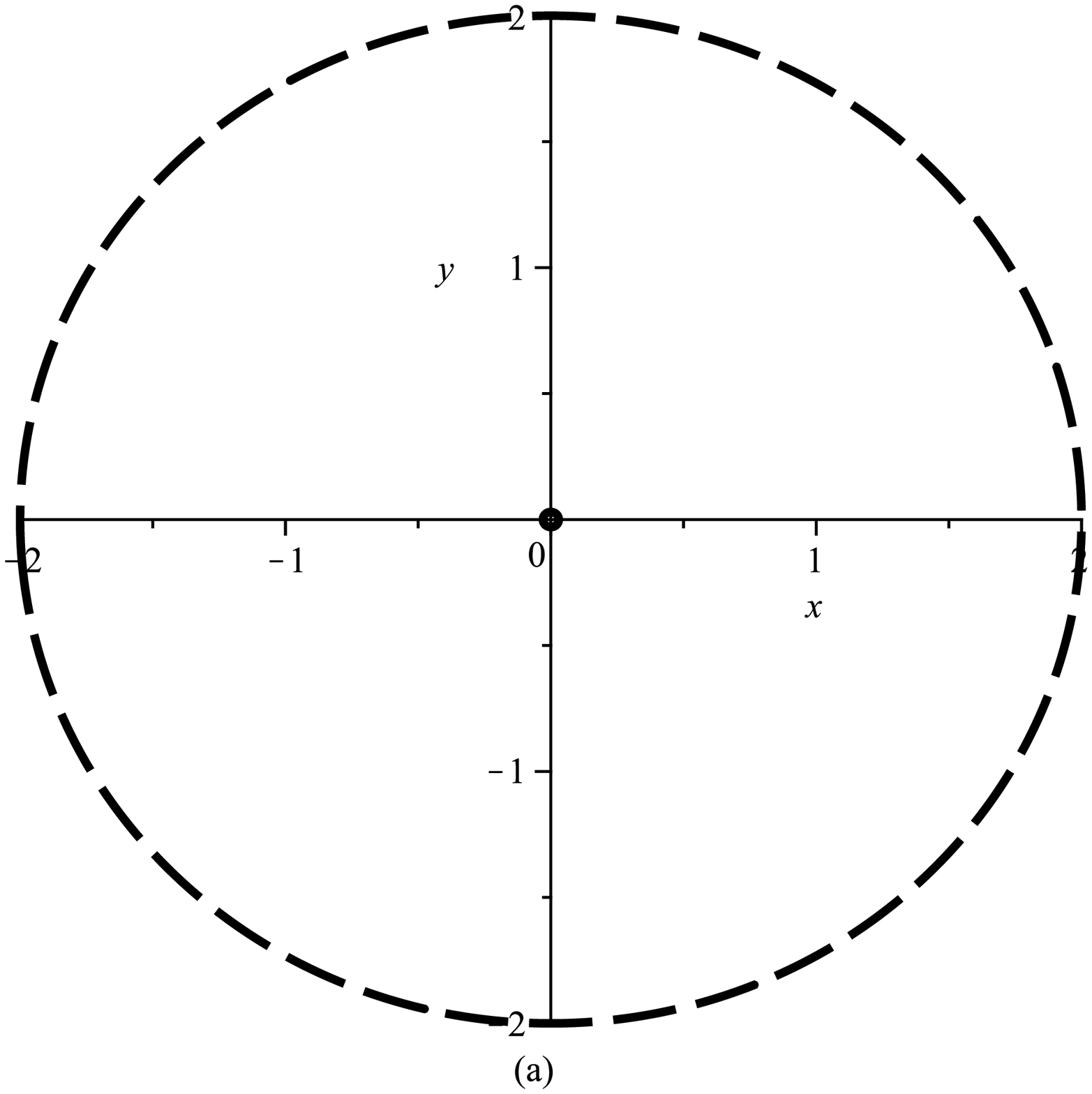}
\includegraphics[scale=0.2]{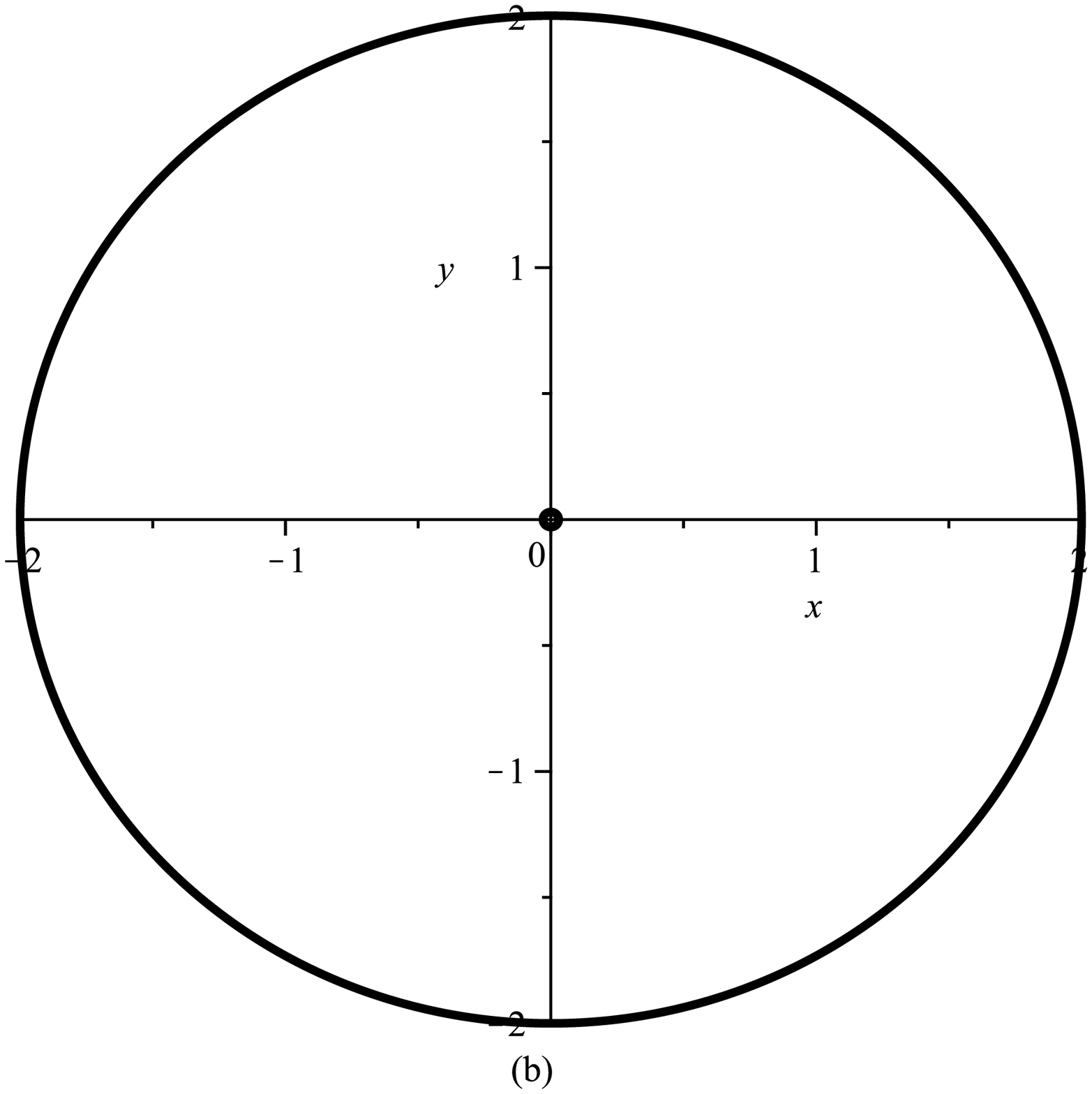}
\includegraphics[scale=0.2]{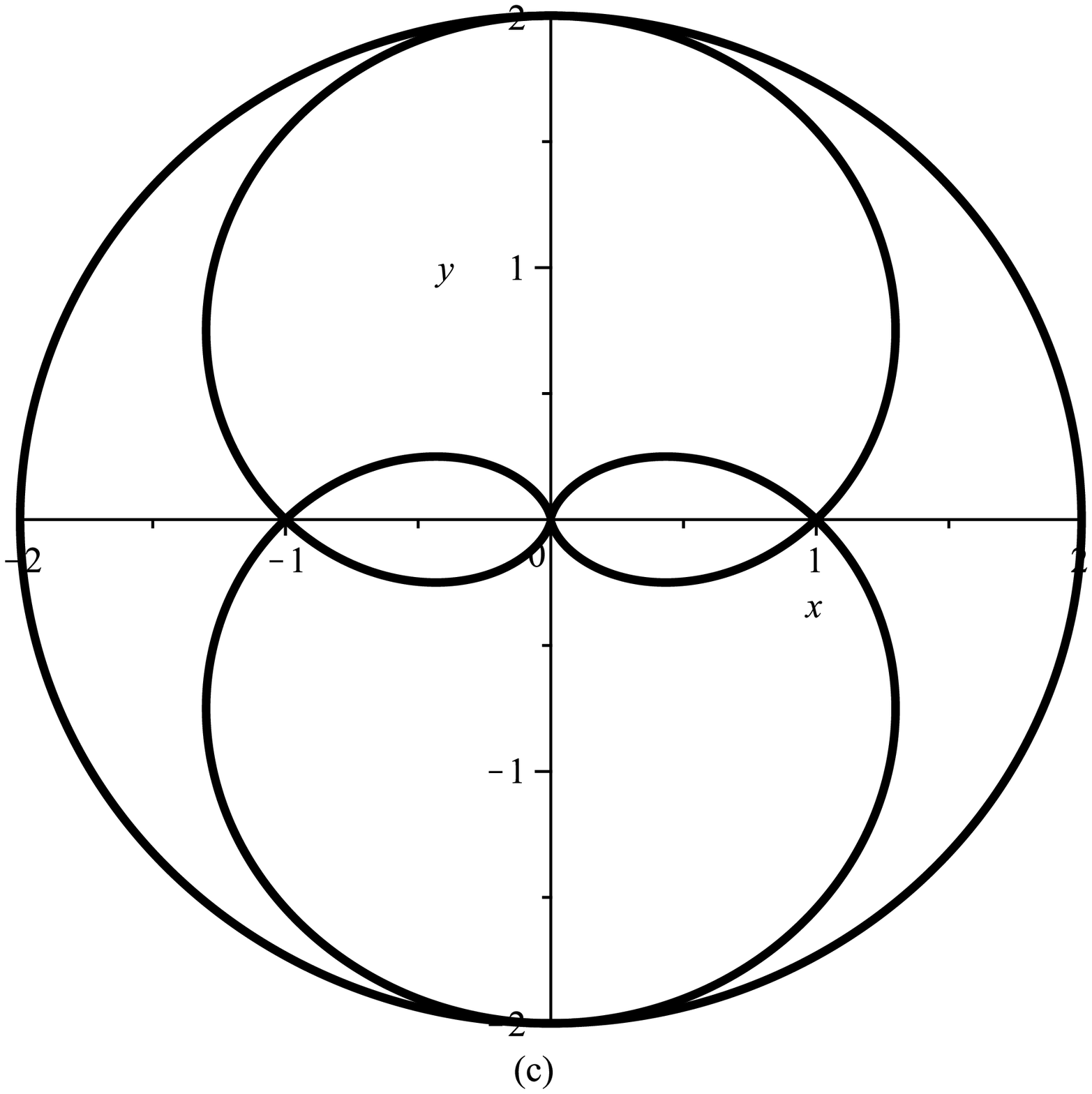}
\caption{Structure of naked singularities of a spacetime with quadrupole. Plot (a) represents the limiting case 
of a Schwarzschild spacetime $(q=0)$ with a singularity at the origin surrounded by the horizon (dashed curve) 
situated at $r=2m$. Once the quadrupole parameter $q$ is included, the horizon transforms into 
a naked singularity (solid curve) and the central singularity becomes naked as well. This case is illustrated in plot (b). 
For values of the quadrupole parameter within the interval $q\in (-1,-1+\sqrt{3/2}]\backslash \{0\}$, two additional
naked singularities appear as depicted in plot (c).}
\label{fig1}
\end{figure}

To investigate the structure of possible curvature singularities, we consider the Kretschmann scalar 
$K = R_{\mu\nu\lambda\tau}R^{\mu\nu\lambda\tau}$. A straightforward computation leads to 
\be
K   = \frac{16 m^2(1+q)^2}{r^{4(2+2q+q^2)}}\frac{ (r^2-2mr+m^2\sin^2\theta)^{2(2q+q^2)-1}}{(1-2m/r)^{2(q^2+q+1)}}L(r,\theta)\ ,
\label{kre}
\ee
with 
\bea
L(r,\theta)= & & 3(r-2m-qm)^2(r^2-2mr+m^2\sin^2\theta) \nonumber\\
& & + q(2+q)\sin^2\theta[ q(2+q) + 3(r-m)(r-2m-qm)] \ .
\eea
In the limiting case $q=0$, we obtain the Schwarzschild value $K= {48 m^2}/{r^6}$ with the only singularity 
situated at the origin of coordinates $r\rightarrow 0$. In general, one can show that the singularity at
the origin is present for any values of $q$. Moreover, an additional singularity appears at the radius $r=2m$ 
which, according to the metric (\ref{zv}), is also a horizon in the sense that the norm of the timelike Killing 
tensor vanishes at that radius. Outside the hypersurface $r=2m$ no additional horizon exists, indicating 
that the singularities situated at the origin and at $r=2m$ are naked. A more detailed analysis of the Kretschmann
scalar  (\ref{kre}) shows that in the interval $q\in (-1,-1+\sqrt{3/2}]\backslash \{0\}$ two further singularities appear 
inside the singular hypersurface $r=2m$. This configuration of naked singularities is schematically illustrated 
in Fig. \ref{fig1}. 
We conclude that the presence of a quadrupole moment completely changes the structure of spacetime.
In particular, naked singularities are always present, regardless  of the value of the quadrupole parameter.  

\section{Geodesic motion}
\label{sec:eff}

\begin{figure}
\begin{center}
\includegraphics[scale=0.35]{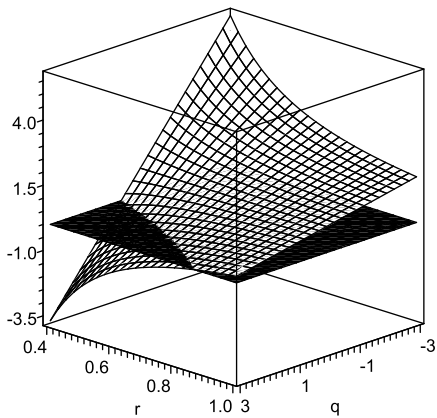}
\caption{Behavior of the effective mass for the metric (\ref{zv}). The effective mass (\ref{meff}) has been evaluated 
on the equatorial plane $\theta=\pi/2$ and is plotted as a function of the parameter $q$ and the radial coordinate $r$.
The plane $M_{eff}=0$ has been plotted to distinguish the region where the effective mass becomes negative. }
\label{fig2}
\end{center}
\end{figure}

The study of the motion of test particles in the gravitational field of naked singularities has shown that 
certain effects appear that can be explained by assuming the existence of an effective gravitational potential
that generates repulsive gravity under specific circumstances. Currently, there is no invariant definition of
repulsive gravity in the context of general relativity, although some attempts have been made by using 
invariant quantities constructed with the curvature of spacetime \cite{def,christian}. Nevertheless, it is possible
to consider an intuitive approach by using the fact that the motion of test particles in stationary axisymmetric
gravitational fields reduces to the motion in an effective potential. This is a consequence of the fact that
the geodesic equations possess two first integrals associated with stationarity and axial symmetry. The explicit 
form of the effective potential depends also on the type of motion under consideration. But in general one can find
certain similitudes between the effective potential for geodesic motion and the effective Newtonian potential which
follows from the metric as $g_{tt} \approx 1 - 2 V_{N} = 1 - 2 M_{eff}/r$. For a mass $M_0$ with quadrupole moment 
$M_2$, the Newtonian potential reads $V_N = M_0/r+(1-3\cos^2\theta)M_2/(4r^3)$ so that in the particular case 
of the metric  (\ref{zv}), we obtain the effective mass
\be
M_{eff} = m(1+q)\left[ 1 -\frac{m^2}{12 r^2} q (2+q) (1-3\cos^2\theta)\right]\ .
\label{meff}
\ee
Clearly, this expression can become negative for certain values of the coordinates and parameters. This behavior 
is illustrated in Fig. \ref{fig2}. 

One can then 
intuitively expect that  in the region of negative effective mass the motion of test particles will experience 
effects that can be interpreted as due to the presence of repulsive gravity. In fact, it is even possible 
to find situations in which repulsive gravity compensates the attractive gravitational force and test 
particles can remain at rest with respect to static observers situated at infinity. We have performed an analysis
of the motion of test particles moving along circular orbits around naked singularities of different types: 
Reissner-Nordstr\"om naked singularity with mass $M$ and electric charge $Q$ with $M<|Q|$, Kerr naked 
singularity with mass $M$ and specific angular momentum $a$ with $M<|a|$, Kerr--Newman naked singularity 
with $M^2<a^2+Q^2$, ZV naked singularity with mass parameter $m$ and arbitrary 
quadrupole parameter $q$, and rotating ZV naked singularity with an additional arbitrary parameter $j$ associated
to the angular momentum. In all these cases we found that it is possible to introduce an effective mass that 
can become negative in a certain region of spacetime and resembles the behavior illustrated in Fig. \ref{fig2}.

\begin{figure}
\begin{center}
\includegraphics[scale=0.7]{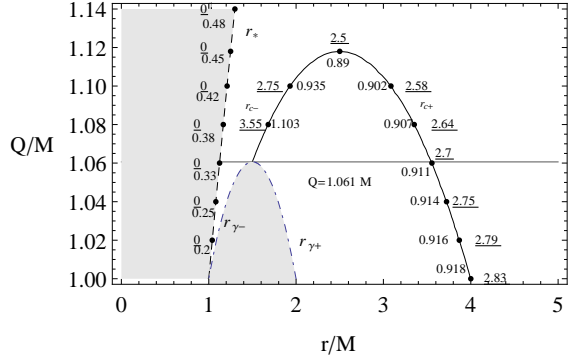}
\caption{The
minimum radius $r_{lsco}/M$ (solid curve) is
plotted as a function of the ratio $Q/M$. Shaded regions are forbidden (only spacelike geodesics are allowed).
The dashed curve represents the classical radius $r_{*}= Q^2/M$ where $L=0$ and the particle stays at ``rest". 
The dotted curve $r_{\gamma^-}$ and the dot-dashed curve $r_{\gamma^+}$ represent lightlike orbits. 
 The region inside the curves $r_{c+}$, $r_{c-}$ and $r_{\gamma^+}$ 
corresponds to the region of instability. Outside this region (and outside the forbidden zone)
any point represents a stable circular orbit. 
Numbers close to the selected points represent the value of the energy $E/\mu$ and the angular
momentum $L/(\mu M)$ (underlined numbers) of the last stable
circular orbits.}  
\label{fig3}
\end{center}
\end{figure}

\begin{figure}
\begin{center}
\includegraphics[scale=0.3]{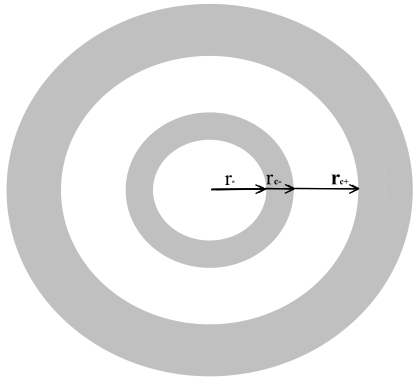}
\caption{Structure of an accretion disk around a naked singularity. The interior ring is situated within the interior region 
of stability.}
\label{fig4}
\end{center}
\end{figure}

A detailed study of the motion of test particles in the gravitational field of naked singularities is beyond the
scope of the present work. In general it is necessary to perform numerical computations to solve the geodesic 
equations. Nevertheless, the case of circular motion around a Reissner-Nordstr\"om naked singularity can be 
studied analytically \cite{pqrmg12} and the results reproduce in general terms those of other naked configurations.
For this reason we present here only the results for the Reissner-Nordstr\"om case. Imagine a test particle of mass
$\mu$ on a circular trajectory around a naked singularity with mass $M$ and charge $Q$. Then, the geodesic motion 
is equivalent to the motion in the effective potential
\begin{equation}
\label{peff}
V_{eff} \equiv\sqrt{\left(1+\frac{L^2}{\mu^2r^2}\right)\left(1-\frac{2M}{r}+\frac{Q^2}{r^2}\right)} \ ,
\end{equation}
where $r$ is the radial coordinate and $L$ the angular momentum of the test particle. It is then possible
to find the radius of the last stable circular orbit $r_{lsco}$. The result is plotted in Fig. \ref{fig3}. 
The most interesting feature is that below the value $Q/M=\sqrt{5}/2$ there are two regions of stability.
Let us suppose that an accretion disk around such a singularity is made off test particles on circular motion.
For the particular value $Q/M=1.1$, particles with radius in the intervals $(r_*,r_{c-})$ and ($r_{c+},+\infty)$ 
are stable. If a particle happens to be within the region of instability  $(r_{c-},r_{c+})$, it necessarily will 
move toward one of the regions of stability (see Fig. \ref{fig3}). Then, the accretion disk will become discontinuous. In a realistic
situation one would expect that the interior ring would be situated outside the classical radius $r_*$ and very close
to the last stable radius $r_{c-}$. The exterior disk would then be situated around the radius $r_{c+}$. This structure 
is schematically illustrated in Fig. \ref{fig4}. A more detailed study of kind of discontinuous accretion disks around naked 
singularities will presented elsewhere.

\section{Conclusions}

In this work we presented an exact solution of Einstein's field equations in empty space which represents
the exterior gravitational field of a static naked singularity. It was obtained by applying a particular 
ZV transformation on the Schwarzschild metric. 
To our knowledge, this is the simplest solution which contains a mass parameter and a quadrupole parameter as well.
The quadrupole moment of the source is identified as the source of the naked singularity. We briefly discuss the motion
of test particles around naked singularities and argue that an effective mass can be introduced that in certain regions
of spacetime can become negative and give rise to repulsive gravity. The analysis of circular motion around naked 
singularities indicates that there exist discontinuous regions of stability. Consequently, a hypothetical accretion
disk made off test particles on circular motion present a discontinuous structure. It would be interesting to 
further analyze this type of accretion disks to determine if they could lead to observational consequences.

In the context of singularities the question about stability is essential. 
It is interesting to notice that only in the case of ZV  naked singularities the parameters that enter the metric 
are arbitrary. This could be interpreted as an indication of the stability of naked singularities generated 
by a quadrupole. In fact, all naked singularities which have a black hole counterpart must satisfy certain 
conditions among the parameters entering the corresponding metric. The analysis of stability indicates that 
perturbations modify these conditions in such a way that the resulting configuration must rapidly decay into
a black hole. In the case of the static ZV naked singularity studied above a perturbation of the parameters $m\rightarrow 
m+\delta m$ and $q\rightarrow q +\delta q$ does not lead to any additional condition among $m$ and $q$. 
A more exhaustive analysis will be necessary to determine if this heuristic argument is indeed an indication 
of stability.

\end{document}